\documentclass[runningheads]{llncs}
\usepackage{amssymb, amsmath, amsfonts}
\usepackage{euscript}
\usepackage{mathtools}
\usepackage{comment}
\usepackage{graphicx}
\usepackage[dvipsnames]{xcolor}
\usepackage{hyperref}
\graphicspath{Figures/}
\usepackage{hyperref}
\usepackage{graphicx}
\usepackage{xcolor}
\usepackage{multirow}
\usepackage{booktabs}
\begin{document}

\title{TransDeepLab: Convolution-Free Transformer-based DeepLab v3+ for Medical Image Segmentation}

\author{Reza Azad\inst{1} \and
Moein Heidari \inst{2} \and Moein Shariatnia \inst{3} \and Ehsan Khodapanah Aghdam \inst{4} \and Sanaz Karimijafarbigloo \inst{1} \and Ehsan Adeli\inst{5} \and Dorit Merhof\inst{1,6}}
\institute{Institute of Imaging and Computer Vision,
RWTH Aachen University, Germany\and
School of Electrical Engineering, Iran University of Science and Technology, Iran
\and Tehran University of Medical Sciences, School of Medicine \and Department of Electrical Engineering, Shahid Beheshti University, Tehran, Iran \and Stanford University \and
Fraunhofer Institute for Digital Medicine MEVIS, Bremen, Germany\\\email{\{dorit.merhof\}@lfb.rwth-aachen.de}}

\authorrunning{Azad, Heidari, Shariatnia, Khodapanah Aghdam, Karimijafarbigloo et al.}
\titlerunning{TransDeepLab: Convolution-Free Transformer-based DeepLab v3+}

\maketitle              

\begin{abstract}
Convolutional neural networks (CNNs) have been the de facto standard in a diverse set of computer vision tasks for many years. Especially, deep neural networks based on seminal architectures such as U-shaped model with skip-connections or atrous convolution with pyramid pooling have been tailored to a wide range of medical image analysis tasks. The main advantage of such architectures is that they are prone to detaining versatile local features. However, as a general consensus, CNNs fail to capture long-range dependencies and spatial correlations due to the intrinsic property of confined receptive field size of convolution operations. Alternatively, Transformer, profiting from global information modeling that stems from the self-attention mechanism, has recently attained remarkable performance in natural language processing and computer vision. Nevertheless, previous studies prove that both local and global features are critical for a deep model in dense prediction, such as segmenting complicated structures with disparate shapes and configurations. To this end, this paper proposes TransDeepLab, a novel DeepLab-like pure Transformer for medical image segmentation. Specifically, we exploit hierarchical Swin-Transformer with shifted windows to extend the DeepLabv3 and model the Atrous Spatial Pyramid Pooling (ASPP) module. A thorough search of the relevant literature yielded that we are the first to model the seminal DeepLab model with a pure Transformer-based model. Extensive experiments on various medical image segmentation tasks verify that our approach performs superior or on par with most contemporary works on an amalgamation of Vision Transformer and CNN-based methods, along with a significant reduction of model complexity. The codes and trained models are publicly available at \href{https://github.com/rezazad68/transdeeplab}{\textcolor{red}{github}.}.

\end{abstract}


\section{Introduction}
Automatic and accurate medical image segmentation, which consists of automated delineation of anatomical structures and other regions of interest (ROIs), plays an integral role in the assessment of computer-aided diagnosis (CAD) \cite{chen2021transunet,li2021medical,heidari2022hiformer,feyjie2020semi,azad2021texture,azad2021stacked}. As a flagship of deep learning, convolutional neural networks (CNNs) have scattered existing contributions in various medical image segmentation tasks for many years \cite{ronneberger2015u,milletari2016v,azad2022medical,azad2022intervertebral,azad2022smu}. Among diverse CNN variants, the widely acknowledged symmetric Encoder-Decoder architecture nomenclature as U-Net \cite{ronneberger2015u} has demonstrated eminent segmentation potential. It mainly consists of a series of continuous convolutional and down-sampling layers to capture contextual semantic information through the contracting path. Then in the decoder, using lateral connections from the encoder, the coarse-grained deep features, and fine-grained shallow feature maps are up-sampled to generate a precise segmentation map. Following this technical route, many U-Net variants such as U-Net++ \cite{zhou2018unet++} and Res-UNet \cite{zhang2018road} have emerged to improve the segmentation performance. A paramount caveat of such architectures is the gap of restricted receptive field size, which makes the deep model unable to capture sufficient contextual information, causing the segmentation to fail in complicated areas such as boundaries. To mitigate this problem, the notable DeepLab \cite{chen2014semantic} work was exhibited, triggering broad interest in the image segmentation era. The authors established remarkable contributions which experimentally proved to have substantial practical merit. First, they introduced a novel convolution operation with up-sampled filters called ‘Atrous Convolution’, which allows enlarging the field of view of filters to absorb larger contexts without imposing the burden of the high amount of computation or increasing number of parameters. Second, to incorporate smoothness terms enabling the network to capture fine details, they exploit a fully connected Conditional Random Field (CRF) to refine the segmentation results. Following the pioneering work, extended versions were employed to accommodate further performance boosts. As such, the DeepLabv2 \cite{chen2017deeplab} was proposed to conquer the challenge of the existence of objects at multiple scales. To this end, they propose the atrous spatial pyramid pooling (ASPP) module to segment objects at multiple scales robustly. ASPP probes a feature map with multiple atrous convolutions with different sampling rates to obtain multi-scale representation information. Afterward, the DeepLabv3 \cite{chen2017rethinking} designed an Encoder-Decoder architecture with atrous convolution to attain sharper object boundaries, where they utilized depth-wise separable convolution to increase computational efficiency. Ultimately, Chen et al. \cite{chen2018encoder} proposed the DeepLabv3+ that extends DeepLabv3 by adding a simple yet effective decoder module to facilitate the segmentation performance. Despite all the efforts, the shortcomings of CNNs are also very prominent as they inevitably have constraints in learning long-range dependency and spatial correlations due to their inductive bias of locality and weight sharing \cite{xie2021cotr} that results in sub-optimal
segmentation of complex structures. Recently, the novel architecture Transformer \cite{vaswani2017attention} has sparked discussions in computer vision era \cite{dosovitskiy2020image,hatamizadeh2022unetr} due to its elegant design and existence of attention mechanism. Indeed, it has been witnessed as capable of learning long-term features and felicitously modeling global information. The pioneering Vision Transformer (ViT) \cite{dosovitskiy2020image} was the major step toward adapting Transformers for vision tasks which accomplished satisfactory results in image classification. It mainly proposed to split the input image into patches and consider them as the source of information for the Transformer module. Despite being feasibly designed, the drawbacks of this scenario are noticeable and profound \cite{cao2021swin}. First, Transformers impose a quadratic computational load, making it intolerable for dense prediction with high-resolution image tasks. Moreover, despite being a good design choice for capturing explicit global context and long-range relations, Transformers are weak in capturing low-level pixel information, which is indisputably crucial in developing accurate segmentation. Thus, to circumvent the high memory demand in Transformers, the Swin-Transformer \cite{liu2021swin} proposed a hierarchical ViT with local computing of self-attention with non-overlapping windows, which achieved a linear complexity as opposed to ViT. Recently, faced with the dilemma between efficient CNNs and powerful ViT, crossovers between the two areas have emerged where most try to model a U-Net-like architecture with Transformers. Examples of such are Trans-UNet \cite{chen2021transunet}, Swin-UNet \cite{cao2021swin}, and DS-TransUNet \cite{lin2021ds}. Inspired by the breakthrough performance of DeepLab models with attention mechanism in segmentation tasks \cite{azad2020attention}, in this paper, we propose TransDeepLab, a DeepLab-like pure Transformer for medical image segmentation. Akin to the recently proposed Swin-UNet that models a U-Net structure with a Transformer module, we aim to imitate the seminal DeepLab with Swin-Transformer. The intuition behind our choice is that we intend to facilitate the efficient deployment of Swin-Transformer to restrain the hinder of computational demand of ViT. Moreover, applying the Swin-Transformer module with multiple window sizes can make it a lightweight yet suitable design choice for multi-scale feature fusion, which is a particularly critical equipment in segmentation tasks. In particular, we aim to substitute the ASPP module of the DeepLabv3+ model with the aforementioned hierarchical design. All these lead us to the fact that the proposed TransDeepLab can be the optimal design that is able to efficiently compensate for the mediocre design flaws of DeepLab. The proposed method acquires a significant parameter decrease compared to the cohort study. We will elaborate on the details of our proposal by pinpointing the scope and contributions of this paper in Section 2. Our contributions are as follows: \textbf{(1)} By incorporating the advantages of hierarchical Swin-Transformer into the encoder, decoder, and ASPP module of DeepLab, the proposed TransDeepLab can effectively capture long-range and multi-scale representation. \textbf{(2)} The cross-contextual attention to adaptively fuse multi-scale representation. \textbf{(3)} To the best of our knowledge, this work is the first attempt to combine the Swin-Transformer with DeepLab
architecture for medical image segmentation.

\section{Proposed method}
\begin{figure}[h]
	\centering
		\includegraphics[width=\textwidth]{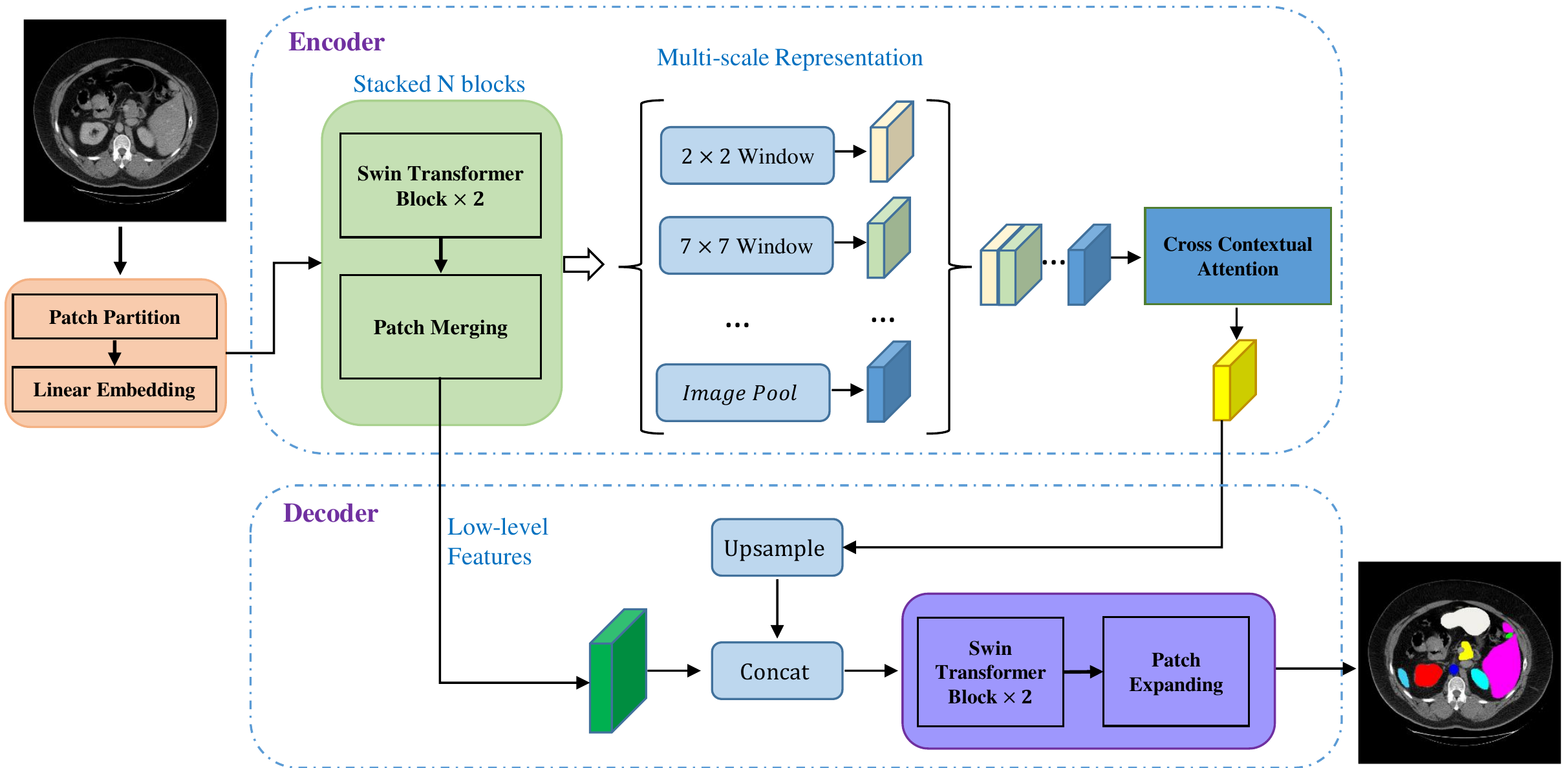}
	\caption{The architecture of TransDeepLab, which extends the encoder-decoder structure of DeepLabv3+. Encoder and decoder are all constructed based on Swin-Transformer blocks.} \label{fig:proposed_method}
\end{figure}

We propose the TransDeepLab model (\autoref{fig:proposed_method}), a pure Transformer-based DeepLabv3+ architecture, for medical image segmentation.  The network utilizes the strength of the Swin-Transformer block \cite{liu2021swin} to build hierarchical representation. Following the original architecture of the DeepLab model, we utilize a series of Swin-Transformer blocks to encode the input image into a high-representational space.  More specifically, the encoder module splits the input medical image into non-overlapping patches of size $4 \times 4$, resulting in $4 \times 4 \times 3=48$ as the feature dimension of each patch (signified as C) and applies the Swin-Transformer block to encode both local semantic and long-range contextual representation. To model Atrous Spatial Pyramid Pooling (ASPP), a pyramid of Swin-Transformer blocks with varying window sizes is designed. The main idea of the Swin pyramid is to capture multi-scale information by exploiting different window sizes. The obtained multi-scale contextual representation is then fused into the decoder module using a Cross-Contextual attention mechanism. The attention block applies two-level attention (e.g., channel and spatial attention) on the tokens (derived from each level of the pyramid) to formulate the multi-scale interaction. Finally, in the decoding path, the extracted multi-scale features are first bilinearly upsampled and then concatenated with the low-level features from the encoder to refine the feature representation. The details of each component of the proposed network will be elaborated on in the subsequent sections.

\subsection{Swin-Transformer block}
Based on the fact that typical vision Transformers implement the self-attention on a global receptive field, they endure quadratic computational complexity to the number of tokens. To mitigate this, the Swin-Transformer has been devised whose key design characteristic is its shifting of the window partitioner between consecutive self-attention layers constructed by designing a module based on shifted windows as a surrogate for the multi-head self-attention (MSA) module in a Transformer block. Thus, a Swin-Transformer block comprises a shifted window-based MSA module, LayerNorm (LN) layer, a two-layer MLP, and GELU nonlinearity. The window-based multi-head self-attention (W-MSA) module and the shifted window-based multi-head self-attention (SW-MSA) module are applied in the Transformer blocks in tandem. With such shifted window partitioning scheme, consecutive Swin-Transformer blocks can be formulated as:
\begin{align}
\hat{z}^{l}&=\text{W-MSA}\left(\text{LN}\left(z^{l-1}\right)\right)+z^{l-1} \nonumber \\
z^{l}&=\text{MLP}\left(\text{LN}\left(\hat{z}^{l}\right)\right)+\hat{z}^{l} \nonumber \\
\hat{z}^{l+1}&=\text{SW-MSA}\left(\text{LN}\left(z^{l}\right)\right)+z^{l} \nonumber \\
z^{l+1}&=\text{MLP}\left(\text{LN}\left(\hat{z}^{l+1}\right)\right)+\hat{z}^{l+1} ,
\end{align}
where $\hat{z}^{l}$ and $z^{l}$ denote the outputs of W-MSA and SW-MSA module of the $l^{t h}$ block, respectively. Following \cite{hu2018relation,hu2019local} the self-attention is computed according to: 

\begin{align}
\text{Attention}(Q, K, V)=\text{SoftMax}\left(\frac{Q K^{T}}{\sqrt{d}}+B\right) V ,
\end{align}
where $Q, K, V \in \mathbb{R}^{M^{2} \times d}$ are the query, key and value matrices; $d$ is the query/key dimension, and $M^{2}$ is the number of patches in a window and $B$ indicates the bias matrix whose values are acquired from  $\hat{B} \in$ $\mathbb{R}^{(2 M-1) \times(2 M+1)}$.

\subsection{Encoder}

Inspired by the low computation burden of the Swin-Transformer \cite{liu2021swin} block (contrary to the quadratic computation of the Vision Transformer \cite{dosovitskiy2020image}) and its strength in modeling long-range contextual dependency (unlike regular CNNs), we model our encoder model using the stacked Swin-Transformer module. Our TransDeepLab encoder first feeds the C-dimensional tokenized input with the resolution of $\frac{H}{4} \times \frac{W}{4}$ into two successive Swin-Transformer blocks to produce a hierarchical representation while keeping the resolution unchanged. Then, it applies a series of stacked Swin-Transformer blocks to gradually reduce the spatial dimension (similar to a CNN encoder) of the feature map and increase the feature dimension. The resulted mid-level representation is then fed to the Swin Spatial Pyramid Pooling (SSPP) block to capture multi-scale representation.

\subsection{Swin Spatial Pyramid Pooling}

The spatial resolution of the deep features extracted by the encoder module is considerably decreased due to the stacked Swin-Transformer blocks followed by the patch merging layers (similar to the consecutive down-sampling operation in a CNN encoder). Thus, to compensate for the spatial representation and produce a multi-scale representation, the DeepLab model utilizes an ASPP module, which replaces the pooling operation with atrous convolutions \cite{chen2017deeplab}. Concretely, DeepLab aims to form a pyramid representation by applying parallel convolution operations with multiple atrous rates. To model such an operation in a pure Transformer fashion, we create a Swin Spatial Pyramid Pooling (SSPP) block with varying window sizes to capture multi-scale representation. In our design, the smaller window size aims to capture local information while the larger windows are included to extract global information. The resulted multi-scale representation is then fed to a cross-contextual attention module to fuse and capture a generic representation in a non-linear technique.

\subsection{Cross-Contexual Attention}
In the DeepLabv3+ model, the feature vectors resulting from each level of the pyramid are concatenated and fed to the depthwise separable convolution to perform the fusion operation. This operation performs the convolution for each channel separately and is thus unable to model the channel-wise dependency among pyramid levels. In our design, to model the multi-scale interaction and fuse the pyramid features, we propose a cross-attention module. To this end, we assume that each level of the pyramid ($\mathbf{z}_{m}^{P \times C}$, $P$ and $C$ indicate the number of token and embedding dimension, respectively) represents the object of interest in different scales, thus, by concatenating all these features in a new dimension we create a multi-scale representation $\mathbf{z}_{all}^{P \times MC}=[\mathbf{z}_{1} \| \mathbf{z}_{\text {2}} ... \| \mathbf{z}_{\text {M}}]$, where $\|$ shows the concatenation operation. Next, to adaptively emphasize the contribution of each feature map and surpass the less discriminative features, we propose a scale attention module. Our attention module takes into account the global representation of each channel and applies the MLP layer to produce the scaling coefficients ($w_{scale}$) to selectively scale the channel representation among pyramid levels:


\begin{align}
w_{\text {scale }}=\sigma\left(\mathbf{W}_{2} \delta\left(\mathbf{W}_{1} G A P_{z_{all}}\right)\right), z_{all }^{\prime}=w_{\text {scale }} \cdot z_{all}
\end{align}
where $W_1$ and $W_2$ indicate the learnable MLP parameters and $\delta$ and $\sigma$ show the ReLU and Sigmoid activations, and the GAP indicates the global average pooling. In the second attention level, we learn scaling parameters to highlight the informative tokens. To do so, we
apply the same strategy: 
\begin{align}
w_{\text {tokens }}=\sigma\left(\mathbf{W}_{3} \delta\left(\mathbf{W}_{4} G A P_{z_{all}^{\prime}}\right)\right), z_{all}^{\prime \prime}=w_{\text {tokens }} \cdot z_{all}^{\prime}
\end{align}

\subsection{Decoder}
In the decoder, the acquired features ($z_{all}^{\prime\prime}$) corresponding to the attention module are first passed through the Swin-Transformer block with a patch-expanding operation to be upsampled by a factor of 4 and then concatenated with the low-level features.
The scheme of concatenating the shallow features and the deep features together helps in reducing the loss of spatial details by the virtue of down-sampling layers. Finally, a series of cascaded Swin-Transformer blocks with path-expanding operations are applied to reach the full resolution of $H \times W$.

\section{Experiments}
\subsection{Datasets}

\subsubsection{Synapse Multi-Organ Segmentation.}
This dataset includes 30 abdominal CT scans with 3779 axial contrast-enhanced clinical images in total. Each CT posess volumes in range of $85 \sim 198$ slices of $512 \times 512$ pixels, with a voxel spatial resolution of $([0.54 \sim 0.54] \times[0.98 \sim 0.98] \times[2.5 \sim 5.0]) 
 \mathrm{mm}^{3}$. 
We follow \cite{chen2021transunet} in data partitioning and reporting the quantitative results.

\subsubsection{Skin Lesion Segmentation.}
Our analysis for skin lesion segmentation was based on the ISIC 2017 \cite{codella2018skin}, ISIC 2018 \cite{codella2019skin} and $\mathrm{PH}^{2}$ \cite{mendoncca2013ph} datasets.
The ISIC datasets were collected by the International Skin Imaging Collaboration(ISIC) as a large-scale dataset of dermoscopy images  along with their corresponding ground truth annotations. Furthermore, we exploit the $\mathrm{PH}^{2}$ dataset and pursue the experimental setting used in \cite{liu2019enhanced} for splitting the data.

\subsection{Implementation Details}
Turning to implementation aspects, the proposed TransDeepLab is implemented based on the PyTorch library and trained on a single Nvidia RTX 3090 GPU. We train all of our models upstream using the SGD solver in 200 epochs using a batch size of 24. The softmax Dice loss and cross-entropy loss are employed as objective functions, and L2 Norm is also adopted for model regularization. Rotation and flipping techniques are used as data augmentation methods with the aim of diversifying the training set and obtaining an unbiased training strategy. An initial learning rate of 0.05 with an adaptive decay value is used to train the model. In addition, we use the pre-trained weights on ImageNet for the Swin-Transformer module to initialize their parameters. We embraced a task-specific approach to the scope of evaluation metrics aiming to trigger a fair comparison with respect to each experiment. These metrics include: 1) Dice Similarity Score, 2) Hausdorff Distance, 3) Sensitivity and Specificity, 4) Accuracy.

\subsection{Evaluation Results}
In this section, we conduct experiments to evaluate our proposed model and compare it with SOTA methods on the two aforementioned medical image segmentation tasks. Notably, we assess TransDeepLab in two distinct ways in our experiments, i.e., quantitative analysis and along with selected visualization results.

\subsubsection{Results of Synapse Multi-Organ Segmentation}
\begin{figure}[h]
	\centering
		\includegraphics[width=0.7\textwidth]{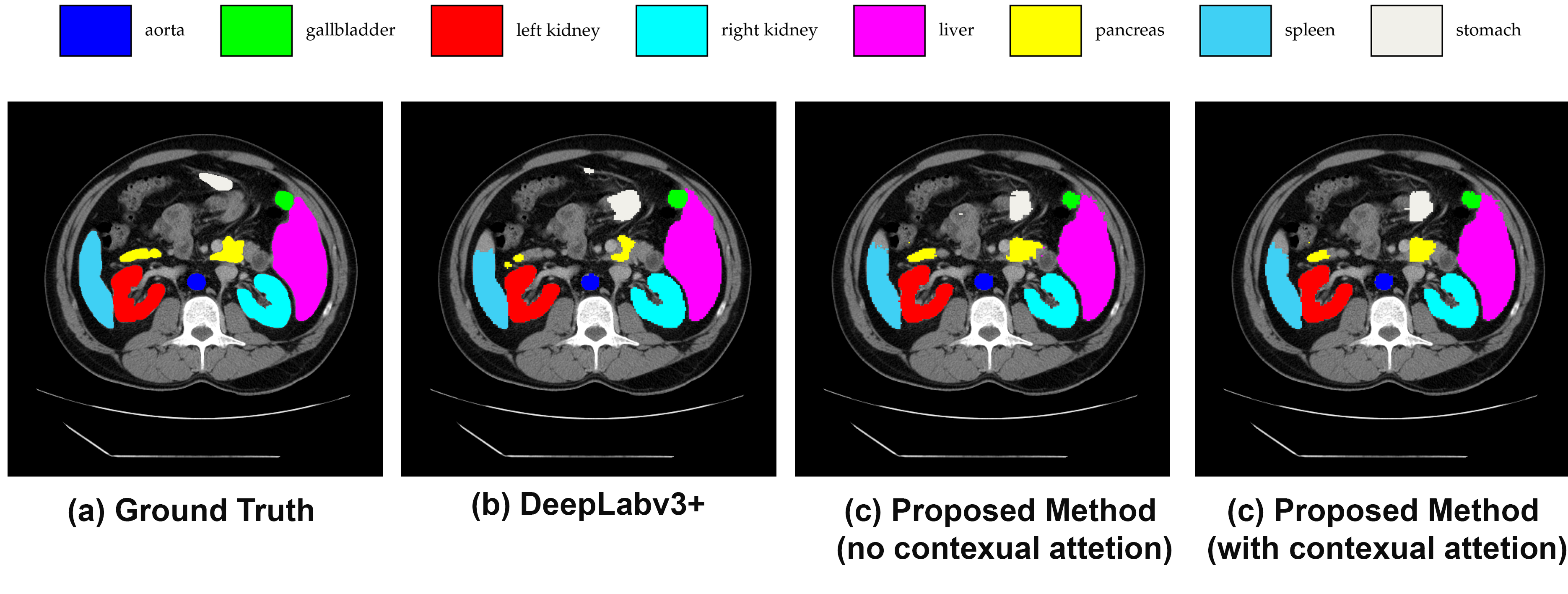}
	\caption{Visualization result of the proposed method on the Synapse dataset.} \label{fig:synapse_vis}
\end{figure}

Experiment on Synapse multi-organ CT dataset (\autoref{tab:results_synapse}) exhibit the effectiveness and generalization potential of our method, achieving the best performance with segmentation accuracy of $80.16 \%(\mathrm{DSC} \uparrow)$ and $21.25 \%(\mathrm{HD} \downarrow)$. Indicatively, we attain the best performance on Kidney(L) with $84.08 \%$, Pancreas with $61.19 \%$, and Stomach with $78.40 \%$ dice score. A sample of segmentation results of synapse multi-organ is presented in \autoref{fig:synapse_vis}. The organ instances are all detected and classified correctly with slight variations in segmentation contours. Compared to the CNN-based DeepLab model, our approach produces better segmentation results. All in all, these results support our ultimate motivation of modeling both local and global contextual representation with a pure Transformer-based method along with providing a significant performance boost in the field of segmentation, where maintaining rich semantic information is crucial.

\begin{table}
\centering
\caption{Comparison results of the proposed method on the Synapse dataset.}
\label{tab:results_synapse}
\resizebox{\columnwidth}{!}{
\begin{tabular}{l|cc|*{8}c}
\hline Methods & DSC~$\uparrow$ &HD~$\downarrow$& Aorta & Gallbladder & Kidney(L) & Kidney(R) & Liver & Pancreas & Spleen & Stomach \\

\hline V-Net \cite{milletari2016v} & $68.81$ & $-$ & $75.34$ & $51.87$ & $77.10$ & $\textbf{80.75}$  & $87.84$ & $40.05$&$80.56$& $56.98$\\

R50 U-Net \cite{chen2021transunet} & $74.68$ & $36.87$ &$87.74$&$63.66$&$80.60$&$78.19$&$93.74$&$56.90$&$85.87$&$74.16$
\\

U-Net \cite{ronneberger2015u} & $76.85$ & $39.70$ &$89.07$&$\textbf{69.72}$&$77.77$&$68.60$&$93.43$&$53.98$&$86.67$&$75.58$
\\

R50 Att-UNet \cite{chen2021transunet} & $75.57$ & $36.97$ &$55.92$&$63.91$&$79.20$&$72.71$&$93.56$&$49.37$&$87.19$&$74.95$
\\

Att-UNet \cite{oktay2018attention} & $77.77$ & $36.02$ &$\textbf{89.55}$&$68.88$&$77.98$&$71.11$&$93.57$&$58.04$&$87.30$&$75.75$
\\

R50 ViT \cite{chen2021transunet}  & $71.29$ & $32.87$ &$73.73$&$55.13$&$75.80$&$72.20$&$91.51$&$45.99$&$81.99$&$73.95$
\\

TransUnet \cite{chen2021transunet} & $77.48$ & $31.69$ &$87.23$&$63.13$&$81.87$&$77.02$&$94.08$&$55.86$&$85.08$&$75.62$
\\

SwinUnet \cite{cao2021swin} & $79.13$ & $21.55$ &$85.47$&$66.53$&$83.28$&$79.61$&$\textbf{94.29}$&$56.58$&$\textbf{90.66}$&$76.60$
\\

DeepLabv3+ (CNN) \cite{chen2017deeplab} & $77.63$ & $39.95$ &$88.04$&$66.51$&$82.76$&$74.21$&$91.23$&$58.32$&$87.43$&$73.53$ 
\\

\hline
\textbf{Proposed Method} & $\textbf{80.16}$ & $\textbf{21.25}$ & $86.04$ & $69.16$ & $\textbf{84.08}$ & $79.88$ & $93.53$ & $\textbf{61.19}$ & $89.00$ & $\textbf{78.40}$
\\
\hline
\end{tabular}
}
\end{table}

\subsubsection{Results of Skin Lesion Segmentation}
\begin{figure}[h]
\centering
\begin{tabular}{@{} *{6}c @{}}
{\tiny Image} & {\tiny Ground Truth} & {\tiny Prediction} & {\tiny Image} & {\tiny Ground Truth} & {\tiny Prediction} \\
\includegraphics[width=0.16\textwidth]{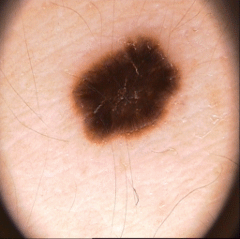} &
\includegraphics[width=0.16\textwidth]{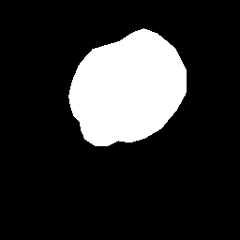} &
    \includegraphics[width=0.16\textwidth]{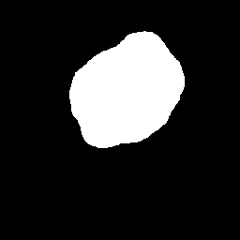} &
\includegraphics[width=0.16\textwidth]{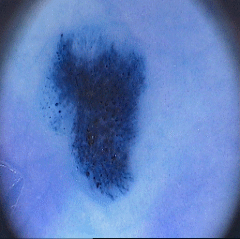} &
\includegraphics[width=0.16\textwidth]{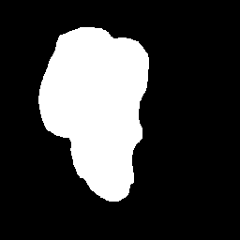} &
\includegraphics[width=0.16\textwidth]{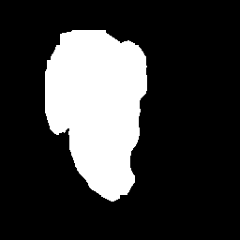} \\
\end{tabular}
\caption{Segmentation results of the proposed method on the skin lesion segmentation.} \label{fig:isic_viz}
\end{figure}

The results are summarized in \autoref{tab:results_isic}. Our TransDeepLab performs better than other competitors w.r.t. most of the evaluation metrics. We also show some samples of the skin lesion segmentation obtained by the suggested network in \autoref{fig:isic_viz}. It is evident from \autoref{fig:isic_viz} that TransDeepLab exhibits higher boundary segmentation accuracy together with a performance boost in capturing the fine-grained details.

\begin{table}[h] 
	\caption{Performance comparison of the proposed method against the SOTA approaches on skin lesion segmentation benchmarks.}\label{tab:results_isic}

	\resizebox{\textwidth}{!}{
		\begin{tabular}{c||c||c||c}
			\hline
			{\begin{tabular}{lccc}
					\multicolumn{4}{l}{\multirow{2}{*}{Methods}} \\
					\textbf{} & \textbf{} & \textbf{}&\textbf{} \\
					\hline
				U-Net~\cite{ronneberger2015u} \\
				Att U-Net~\cite{oktay2018attention}\\
				DAGAN~\cite{lei2020skin}\\
				TransUNet~\cite{chen2021transunet}\\
				MCGU-Net~\cite{asadi2020multi}  \\
				MedT~\cite{valanarasu2021medical}\\
				FAT-Net~\cite{wu2022fat}\\
				TMU-Net \cite{reza2022contextual}\\
				Swin\,U-Net \cite{cao2021swin}\\
				DeepLabv3+ (CNN) \cite{chen2017deeplab}\\
				\hline
				\multicolumn{4}{l}{\textbf{Proposed Method}} \\
				\hline
	
				\end{tabular}
			} &
			{\begin{tabular}{cccc}
					\multicolumn{4}{c}{\textbf{ISIC 2017}} \\
					\hline
					\textbf{DSC} & \textbf{SE} & \textbf{SP}&\textbf{ACC} \\
					\hline
                     0.8159 & 0.8172 & 0.9680 & 0.9164\\
                     0.8082 & 0.7998 & 0.9776 & 0.9145\\
                     0.8425 & 0.8363 & 0.9716 & 0.9304\\
                     0.8123&0.8263&0.9577&0.9207\\
					 0.8927 & 0.8502 & 0.9855 &  0.9570\\
					 0.8037 & 0.8064 & 0.9546 & 0.9090\\
					 0.8500 & 0.8392 & 0.9725 & 0.9326\\
					 0.9164 & \textbf{0.9128} & 0.9789 &  0.9660\\
					 0.9183    & 0.9142    & 0.9798    &  0.9701   \\
					 0.9162    & 0.8733    & \textbf{0.9921}    &  0.9691   \\
					\hline
					\textbf{0.9239}    & 0.8971    & 0.9886    &  \textbf{0.9705}   \\
					\hline

				\end{tabular}
			} &
			{\begin{tabular}{cccc}
					\multicolumn{4}{c}{\textbf{ISIC 2018}} \\
					\hline
					\textbf{DSC} & \textbf{SE} & \textbf{SP} & \textbf{ACC} \\
					\hline
		             0.8545 & 0.8800 & 0.9697 &  0.9404  \\
					 0.8566 & 0.8674 & 0.9863 & 0.9376 \\
				     0.8807 & 0.9072 & 0.9588 & 0.9324 \\
			         0.8499 & 0.8578 & 0.9653 & 0.9452\\
				     0.895 & 0.848 & 0.986 & 0.955 \\
				     0.8389 & 0.8252 & 0.9637 & 0.9358\\
				     0.8903 & \textbf{0.9100} & 0.9699 & 0.9578\\
				     0.9059 & 0.9038 & 0.9746 & 0.9603\\
				     0.8946    & 0.9056    & 0.9798   &  0.9645  \\
				     0.882    & 0.856    & 0.977    &  0.951   \\
				    \hline
				     \textbf{0.9122}    & 0.8756    & \textbf{0.9889}    &  \textbf{0.9654}   \\
				   \hline

				\end{tabular}
			} &
			{\begin{tabular}{cccc}
					\multicolumn{4}{c}{\textbf{PH2}} \\
					\hline
					\textbf{DSC} & \textbf{SE} & \textbf{SP} & \textbf{ACC} \\
					\hline
		             0.8936 & 0.9125 & 0.9588 & 0.9233\\
		             0.9003 & 0.9205 & 0.9640 & 0.9276\\
	                 0.9201&0.8320&0.9640&0.9425\\
	                 0.8840&0.9063&0.9427&0.9200\\					 
					 0.9263 & 0.8322 & 0.9714  & 0.9537\\
					 0.9122 & 0.8472 & 0.9657  & 0.9416\\
					 0.9440 & \textbf{0.9441} & 0.9741 & \textbf{0.9703}\\
					 0.9414 & 0.9395 & 0.9756 & 0.9647\\
					 0.9449    & 0.9410    & 0.9564    &  0.9678   \\
					 0.9202    & 0.8818    & 0.9832    &  0.9503   \\
					\hline
					\textbf{0.9456}    & 0.9161    & \textbf{0.9896}    &  0.9657   \\
					\hline
				\end{tabular}
			} \\
		\end{tabular}
		}
\end{table}
\subsubsection{Model Complexity}Last but not least, we analyze the training parameters of the proposal, as heavy deep nets with small medical image datasets are usually prone to overfitting. TransDeepLab is essentially a lightweight model with only 21.14M parameters. Compared with Swin-UNet \cite{cao2021swin}, the original DeepLab model \cite{chen2017deeplab}, and Trans-UNet \cite{chen2021transunet} which have 27.17M, 54.70M, and 105M parameters respectively, our lightweight TransDeepLab shows great superiority in terms of model complexity whilest being dominant in terms of evaluation metrics.

\subsection{Ablation study}

\subsubsection{CNN vs Transformer Encoder.}
The ablation experiment is conducted to explore the Transformer’s replacement design. In particular, we employed the same decoder and SSPP module as our baseline, but replaced the encoder with a CNN backbone (e.g., ResNet-50) model (denoted as \emph{CNN as Encoder} in \autoref{tab:ablation_study}). Judging from the results of \autoref{tab:ablation_study}, we perceive that a solitary CNN-based encoder yields a sub-optimal performance. Literally, the Transformer module indeed helps TransDeepLab to do segmentation to a certain degree.


\subsubsection{Attention strategy.}
Then, we compared the policy of fusing each level of the Swin-Transformer resulting in multi-scale representation. Concretely, we compare the proposed cross-attention module with a basic scale fusion method, concatenating the feature maps and applying a fully connected layer to fuse them (denoted as \emph{Basic Scale Fusion} in \autoref{tab:ablation_study}). Judging from \autoref{tab:ablation_study}, we deduce that the cross-attention module confirms our intuition of capturing the interaction of feature levels in terms of informativeness of the tokens in different scales.  
Moreover, as for perceptual realism, we have provided sample segmentation results in \autoref{fig:synapse_vis} which indicate that by using the cross contextual attention mechanism we attain closer to the ground truth results, in line with the real situation. This visualization divulges the effect of a multi-scale Transformer module for long-range contextual dependency learning leading to precise localization abilities, especially in boundary areas, a substantial caveat in the image segmentation.


\subsubsection{SSPP influence.}
As discussed above, the SSPP module improves the representation ability of the model in context patterning by probing features at multiple scales to attain multi-scale information. We conduct an inquiry into the feature aggregation from adjacent layers of Swin-Transformer assembling the SSPP module with four sets of combinations which explicitly range from 1 to 4 in our experiments. In \autoref{tab:ablation_study} by comparing the results, we can deduce that using a two-level SSPP module mostly leads to dice score performance gains as it assists in handling scale variability in medical image segmentation. Moreover, we perceive that a three-level SSPP module brings along a notable performance in terms of Hausdorff distance. However, to attain more efficiency, the resolution of the input image should be in compliance with the SSPP level, signifying that increasing the number of SSPP levels should follow a higher resolution image. The results also corroborate the propensity of Transformer in incorporating global context information into the model than its CNN counterpart. While one might speculate that thoroughly modeling a CNN-based network using Transformer would cause model complexity, it is worth mentioning that we aim to overcome this issue by exploiting the Swin-Transformer instead of a typical ViT.

\begin{table}
\centering
\caption{Ablation study on the impact of modifying modules inside the proposed method. We report our results using the Synapse dataset.} \label{tab:ablation_study}
\resizebox{\columnwidth}{!}{
\begin{tabular}{l|cc|cccccccc}
\hline \textbf{Setting} & DSC~$\uparrow$ &HD~$\downarrow$& Aorta & Gallbladder & Kidney(L) & Kidney(R) & Liver & Pancreas & Spleen & Stomach \\ \hline 
CNN as Encoder & $75.89$ & $28.87$ & $85.03$ & $65.17$ & $80.18$ & $76.38$  & $90.49$ & $57.29$ & $85.68$ & $69.93$\\ \hline 
Basic Scale Fusion & $79.16$ & $22.14$ & $85.44$ & $68.05$ & $82.77$ & $80.79$ & $93.80$ & $58.74$ & $87.78$ & $75.96$ \\ \hline 

SSPP Level 1 & $79.01$ & $26.63$ & $85.61$ & $68.47$ & $82.43$ & $78.02$ & $94.19$ & $58.52$ & $88.34$ & $76.46$\\  

SSPP Level 2 & $80.16$ & $21.25$ & $86.04$ & $69.16$ & $84.08$ & $79.88$ & $93.53$ & $61.19$ & $89.00$ & $78.40$\\ 

SSPP Level 3 & $79.87$ & $18.93$ & $86.34$ & $66.41$ & $84.13$ & $82.40$ & $93.73$ & $59.28$ & $89.66$ & $76.99$\\ 

SSPP Level 4 & $79.85$ & $25.69$ & $85.64$ & $69.36$ & $82.93$ & $81.25$ & $93.09$ & $63.18$ & $87.80$ & $75.56$\\

\hline
\end{tabular}
}
\end{table}



\section{Conclusion}

In this paper, we present TransDeepLab, a pure Transformer-based architecture for medical image segmentation. Specifically, we model the encoder-decoder DeepLabv3+ model and leverage the potential of Transformers by using the Swin-Transformer as the fundamental component of the architecture. Showcased on a variety of medical image segmentation tasks, TransDeepLab has shown the potential
to effectively build long-range dependencies and outperforms other SOTA Vision Transformers in our experiments.


\bibliographystyle{splncs04}
\bibliography{Ref}

\end{document}